\documentclass[a4paper,10pt,twocolumn,aps,showpacs,prb]{revtex4-1}
\usepackage{graphicx}
\usepackage{dcolumn}
\usepackage{bm}
\usepackage{slashed}
\usepackage{amssymb}
\usepackage{amsmath}
\usepackage{amsthm}
\usepackage{color}
\usepackage{esint}

\begin{document}

\title{Fractional Topological Insulators: from sliding Luttinger Liquids to Chern-Simons theory}

\author{Raul A. Santos$^{1,2}$, Chia-Wei Huang$^{3}$ Yuval Gefen$^{2}$,  and D.B. Gutman$^{1}$}
\affiliation{$^{1}$Department of Physics, Bar-Ilan University, Ramat Gan, 52900, Israel}
\affiliation{$^{2}$Department of Condensed Matter Physics, Weizmann Institute of Science, Rehovot 76100, Israel}
\affiliation{$^{3}$Max Planck Institute for Solid State Research, Stuttgart, Germany}


\begin{abstract}
The  sliding Luttinger liquid approach is applied  to study fractional topological insulators (FTIs). 
We show that FTI is the low energy fixed point of the theory for  
realistic spin-orbit and electron-electron interactions. 
We find that the topological phase pertains
in the presence of  interaction that breaks the spin invariance 
and its boundaries are even extended by those terms.
Finally we show that one dimensional chiral anomaly  
in the Luttinger liquid leads to the emergence of topological Chern-Simons terms in  the effective gauge theory
of the FTI state.
\end{abstract}

\pacs{}
\preprint{}

\maketitle

\section{Introduction}
Topological insulators (TIs) are materials  that exhibit a spectral gap in the bulk 
and at the same time   have  gapless excitations on their surface.  
The most famous example of a TI is the quantum Hall effect, observed in two dimensional conductors subjected to 
a perpendicular magnetic field at particular filling fractions\cite{Laughlin1981,Halperin1982}.
Recently, another type of TIs, symmetric under time reversal (TR) became a focus  
of  experimental and  theoretical research \cite{Hasan2010, Qi2011}. 
The experimental observation of these TR symmetric TIs has been reported in two 
dimensional\cite{Konig2007,Konig2008,Park2015} and three dimensional materials\cite{Hsieh2008,Hsieh2009,Xia2009,Hsieh2009b,Zhang2009}.

For non-interacting fermionic TIs and topological superconductors (TSs), the full classification 
based on their symmetry class and spatial dimension  has been developed \cite{Kitaev2009,Schnyder2008,Shinsei2010}. 
Among TIs, the simplest class of fermionic insulators invariant under time reversal $\mathcal{T}$ is the AII 
(or symplectic) class, which is described by a $\mathbb{Z}_2$ topological invariant in 2+1 dimensions. 
A topologically non-trivial state within this class possess a helical edge state \cite{Kane2005}.

For interacting systems the classification of TIs \cite{Ludwig2012,Levin2012}, 
as well as their microscopic description, is a subject of an active research.
A toy model construction belonging to this class is a system of two non-interacting 
quantum Hall bars subjected to opposite magnetic fields, placed one on top of the other. 
The ground state is then composed of two species of electrons with opposite spin polarizations. 
As was pointed  out in Ref \onlinecite{Levin2009}, in a properly tuned state an inclusion of 
electron-electron interactions within each layer leads to two copies of fractional quantum Hall states 
with opposite fillings. This induces the formation of a fractional TI (FTI) state\cite{Bernevig2006,Neupert2011,LSantos2011}, 
stable with respect to time reversal symmetric perturbations \cite{Levin2009}.
Although this model correctly captures the topological properties of the system, several 
simplifying assumptions were made:
(a) The interactions between electrons in the different layers in the bulk was neglected, 
but taken into account between the gapless edge modes.
(b) The spin-orbit interaction was assumed to be $S_z$ conserving  and 
was chosen to correspond to a constant spin-dependent \textquotedblleft magnetic field".

A convenient way of describing TIs and to account for interactions is the sliding Luttinger liquid (LL) approach \cite{Kane2002,Teo2014,Klinovaja2014c}.
Within this framework interactions between parallel Luttinger liquids are engineered to describe different nontrivial phases
in two dimensions. Despite its somewhat artificial appearance, the model 
correctly captures the topological properties of different systems and correctly reproduces the
tenfold classification of noninteracting Hamiltonians \cite{Neupert2014}. 


{\it Connection with previous works.-} Within the sliding LL approach, the construction of the integer and fractional TIs
was carried out in Ref. \onlinecite{Neupert2014} and in Ref. \onlinecite{Klinovaja2014b}. In Ref. \onlinecite{Neupert2014}
the authors construct the different classes of topological systems based on antiunitary symmetries. This was done without
stating any microscopic realization of the Hamiltonian that realizes the symmetries in question. In Ref. 
\onlinecite{Klinovaja2014b} Klinovaja and Tserkovnyak suggest, for the first time in this context of sliding LL approach,
a Hamiltonian with spin orbit coupling that possesses TR symmetry.


In that work the following assumptions were made:
a) spin-orbit interaction was described by  Rashba term, with the $S_z$ component only.
b) The interaction between  the electrons was assumed to be sufficiently strong to stabilize the 
desired fixed point, but was not analyzed explicitly.

In our work we use sliding LL approach, explicitly taking into account generic electron-electron 
interaction terms consistent with TR symmetry in a model microscopic Hamiltonian. 
We also allow for a spin-orbit interaction that acts on all the components 
perpendicular to the wire direction ($z$ and $y$). 
We analyze the stability of the FTI fixed point by deriving and explicitly solving the corresponding RG equations.  
We use this model to obtain an effective Chern-Simon theory as the low energy description and show that it correctly 
captures the low energy physics of Abelian FTI states. Although the connection between Chern-Simons theory and the edge modes
of topological systems is well known \cite{WenBook2004,Ludwig2012}, our work is the first that links
the sliding LL approach with a Chern-Simons low energy description.


This paper is organized as follows: In the first section we review the wire construction of fractional (Abelian) quantum 
Hall states. We then extend the analysis to account for realistic Rashba and electron interactions.
In the second section, we study the relevance of the multi-particle hopping operators that drive the system into the 
topological nontrivial state. We find that the interlayer interaction  makes these operators 
more relevant, in comparison with the toy model limit.  This generic spin orbit interaction is limited by the condition
that it does not close the gap, which would lead  to a transition to a different state, (e.g. through the appearance of a 
nontrivial spin texture).
Finally,  we discuss the emergence of the low energy description that follows from gauge invariance in the problem.
Integrating out the massive modes we show that the low energy model is indeed given by abelian Chern-Simons theory, as 
expected for a FTI state.

\section{FTI from coupled quantum wires}\label{model}
\subsection{Laughlin states from Luttinger Liquids}\label{sec:FQHS}
In this section we briefly review the coupled wire construction developed in Refs \onlinecite{Kane2002,Teo2014}
for classifying  topological states of two dimensional electrons.
This approach  was recently  applied for  anomalous quantum Hall effect \cite{Klinovaja2015}, the Halperin states in the FQHE \cite{Sela2014}, and for the construction of non-Abelian states in FTIs
\cite{Vishwanath2014,Vaezi2014,Sagi2014,Klinovaja2014}. 
We start with an array of parallel identical uncoupled wires separated by a distance $d$. In each wire we place the same 
density $n_e=k^0/\pi d$ of spinless fermions with single particle dispersion $E(k)$. These fermions are subject to a 
magnetic field perpendicular to the plane formed by the wires. We choose the gauge $\vec{A}=-By\hat{x}$ for the vector 
potential. The magnetic field shifts the Fermi momentum in each wire by the amount $\delta k_{F,j}=bj$, with $b=|e|dB/\hbar$. 
The linearized low energy Hamiltonian, around the Fermi momentum $k_{F,j}^\eta=\eta k^0+bj$ for each chirality 
$(\eta=(R,L)=(+,-))$ is

\begin{equation}
 \mathcal{H}_0=v_F^0\sum_{j,\eta}\int dx \eta\psi^{\dagger j}_\eta(-i\partial_x-k_{F,j}^\eta) \psi^j_\eta.
\end{equation}

Under bosonization \cite{GiamarchiBook2003,Coleman1975} (see \ref{app:bosonization}), this Hamiltonian becomes

\begin{equation}
  \mathcal{H}_0=\frac{v^0_F}{2\pi}\sum_j\int dx [(\partial_x\theta^j)^2+(\partial_x\varphi^j)^2],
\end{equation}

\noindent where $\partial\theta_j/\pi$ and $\varphi_j$ are the density and phase fields at wire $j$ respectively. 
Two body density interactions are described by the Hamiltonian 

\begin{equation}
 \mathcal{H}_{{\rm FS}}=\frac{v^0_F}{2\pi}\sum_{jk,\eta\eta'}\int dx\psi^{\dagger j}_\eta\psi^{j}_\eta V_{\eta\eta'}^{jk}\psi^{\dagger k}_{\eta'}\psi^{k}_{\eta'}.
\end{equation}

\noindent After bosonization, the sliding LL Hamiltonian $\mathcal{H}_{\rm SLL}=\mathcal{H}_0+\mathcal{H}_{{\rm FS}}$ takes the 
general (still quadratic) form

\begin{equation}\label{H_sll}
  \mathcal{H}_{\rm SLL}=\frac{v^0_F}{2\pi}\sum_{jk}\int dx (\partial_x{{\phi}_j})^TM_{jk}(\partial_x{\phi}_k),
\end{equation}

\noindent where ${\phi}^T_j=(\varphi_j\,,\,\theta_j)$. The $2\times2$ forward scattering matrix is 
$M_{jk}=\mathbb{I}\delta_{ij}+\mathbb{V}_{jk}$, with $\mathbb{V}_{jk}$ parameterizing the forward scattering interactions.
At the filling fraction 

\begin{equation}\label{filling_frac}
\nu\equiv2k^0/b=1/m,
\end{equation}

\noindent momentum conservation allows for the construction of an infinite set of inter-wire many-particle tunneling operators  
without fast oscillating terms (Friedel oscillations). 
Among these operators, the most relevant - in terms of renormalization group (RG) analysis - is of the form
\begin{equation}\label{tunneling_Laughlin}
 \mathcal{O}_j=\exp[\varphi_j-\varphi_{j+1}+m(\theta_j+\theta_{j+1})],
\end{equation}

\noindent which hops electrons between wires $j$ and $j+1$. In presence of this tunneling term it is convenient
to define the so called link fields \cite{Teo2014}

\begin{eqnarray}\label{link_fields}
 2\bar{\varphi}_{j+\frac{1}{2}}&\equiv&\varphi_j+\varphi_{j+1}+m(\theta_j-\theta_{j+1}),\\
 2\bar{\theta}_{j+\frac{1}{2}}&\equiv&\varphi_j-\varphi_{j+1}+m(\theta_j+\theta_{j+1}).
\end{eqnarray}

\noindent Using the commutation relations between the density and phase fields (\ref{CR}) it's easy to see that the links fields 
satisfy $[\partial_x\bar{\varphi}_{\ell}(x),\bar{\varphi}_{\ell'}(x')]=[\partial_x\bar{\theta}_{\ell}(x),\bar{\theta}_{\ell'}(x')]=0$
and

\begin{equation}
 [\partial_x\bar{\theta}_{\ell}(x),\bar{\varphi}_{\ell'}(x')]=i\pi m\delta_{\ell\ell'}\delta(x-x').
\end{equation}

In the link fields basis, the Hamiltonian becomes

\begin{equation}
 \mathcal{H}=\mathcal{H}_{SLL}+\sum_\ell\int dx g\cos(2\bar{\theta}_\ell).
\end{equation}

When the cosine term is relevant under RG analysis, it opens a gap in the spectrum. This operator can always be made
relevant by putting an appropriate choice of a forward scattering interaction. 
In this state, the system possess an excitation gap and quasiparticles characterizing a Laughlin state at filling $\nu=1/m$.

\subsection{FTI for odd $m$ integers}

\begin{figure}
\includegraphics[width=0.8\columnwidth]{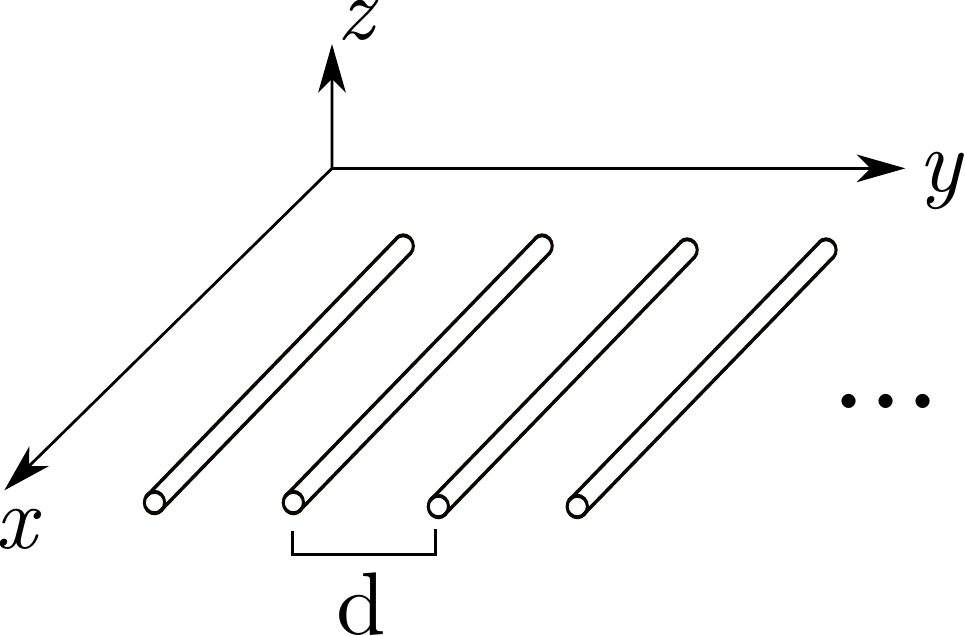}
\caption{\label{fig:wires} Wire arrangement considered for the construction of fractional Hall states and
FTIs with time reversal. In the former case a constant magnetic field $B\hat{z}$ is assumed.
For the FTI construction, the presence of Rashba interaction and a special external confining potential is assumed 
(see text).}
\end{figure}

To construct the TI state we consider spin-orbit interaction in each wires.
The most general non interacting Hamiltonian on each wire $j$, quadratic in momentum including Rashba terms, confining potential and translational invariance 
along the direction of the wire (Fig. \ref{fig:wires}) is

\begin{eqnarray}\label{Ham_Rashba}\nonumber
 H_j&=&\frac{\hat{p}_x^2}{2m_e}+\alpha_{SO}(\vec{p}\times\vec{\sigma})\cdot\nabla V(y,z)+ V(y,z)\\
    &=&\frac{\hat{p}_x^2}{2m_e}+(\lambda^z_j\sigma_z+\lambda^y_j\sigma_y)\hat{p}_x+V_j,
\end{eqnarray}

\noindent where $m_e$ is the electron's effective mass, $\alpha_{SO}$ is the strength of the Rashba coupling and $V_j=V(jd,0)$ is 
a spatially dependent confining potential at the position of each wire (here we assume that the plane formed by the wires is located
at $z=0$). The parameters $\lambda_j^{y,z}$ are simply

\begin{equation}
 \lambda_j^y=\alpha_{SO}\frac{\partial V}{\partial z}\Big|_{\substack{y=jd\\z=0}},\quad \lambda_j^z=-\alpha_{SO}\frac{\partial V}{\partial y}\Big|_{\substack{y=jd\\z=0}}.
\end{equation}

The simplest potential $V$ that leads to the topological nontrivial phase corresponds 
to a parabolic confining potential $V(y,z)=v_1y^2/2+v_2yz$. This potential generates a space dependent
Rashba coupling with strengths $\lambda_j^z=-(\alpha_{SO}v_1d)j$ and $\lambda_j^y=(\alpha_{SO}v_2d)j$
(compare with Ref. \onlinecite{Kane2005} and Eq. (3) in Ref. \onlinecite{Klinovaja2014b}).

The dispersion relation $E_j(k)$ for the Hamiltonian (\ref{Ham_Rashba}) is (with $\hbar=1$)

\begin{equation}
 E_j(k)=\frac{k^2}{2m}+V_j\pm k\sqrt{(\lambda^z_j)^2+(\lambda^y_j)^2}, 
\end{equation}

The eigenstates of (\ref{Ham_Rashba}) are

\begin{eqnarray}
 \psi_{j,+}(x)= e^{ikx}\left(
\begin{array}{c}
 i\sin\alpha_{j}\\\label{eigenstate1}
 \cos\alpha_{j}
\end{array}
\right),\\
\psi_{j,-}(x)=e^{ikx}\left(
\begin{array}{c}
 \cos\alpha_{j}\\\label{eigenstate2}
 -i\sin\alpha_{j}
\end{array}
\right)
\end{eqnarray}

\noindent with $\tan2\alpha_{j}=-\frac{\lambda^y_j}{\lambda^z_j}$. Due to the time reversal symmetry
of the Hamiltonian (\ref{Ham_Rashba}), $\mathcal{T}\psi_{j,s}$ is also an eigenstate with the same energy (Kramers partners).

Note that each Kramers pair is defined up to a phase. This signals the explicit the break of $SU(2)$ invariance (due to the spin-orbit
coupling) to a $U(1)\times U(1)$ symmetry.

To reach the topological phase (see also the discussion after eq. (\ref{mom_cons})), we tune the confining 
potentials such that 

\begin{equation}\label{confining}
v_1=\frac{\Lambda^2}{d^2m_e},\quad \mbox{and} \quad v_2=\frac{\Lambda}{d\alpha_{SO}}\sqrt{1-\left(\frac{\alpha_{SO}\Lambda}{dm_e}\right)^2}.
\end{equation}

The energy dispersion becomes in this case $E^{\pm}(k)=(k\pm j\Lambda)^2/2m_e$. In the case of a single quantum
wire, this energy dispersion can be obtained also by a combination of spin orbit and a Zeeman field. In such scenario
it has been shown  \cite{Oreg2014a} that the system possess fractional quantized conductance and can host 
Majorana bound states by proximity-coupling with a superconductor.

With the choice (\ref{confining}), the Fermi momentum in the wire $j$ becomes 
\begin{equation}\label{Fermi_mom}
 k^{\eta}_{F,s}(j)=\eta k^0-s\Lambda_j,
\end{equation}

\noindent where $\Lambda_j=j\Lambda$ and $\eta=(R,L)=(+,-)$ denotes chirality. In the previous formulas the values 
of $\eta =R/L$ are understood as $\pm1$. The values of $s=(+,-)$ corresponds to the different spinors.  For convenience 
we will refer to electrons with $s=1$ $(-1)$ as those that belong to the upper (lower) layer. 
This choice of potential results in a particularly simple 
dependence of the Fermi momentum on the wire index. This will enable us 
to construct relevant multi-particle tunneling operators that conserve momentum and are free of oscillations,
analogously to how it was done in the Section \ref{sec:FQHS}.

\begin{figure}
\includegraphics[width=1\columnwidth]{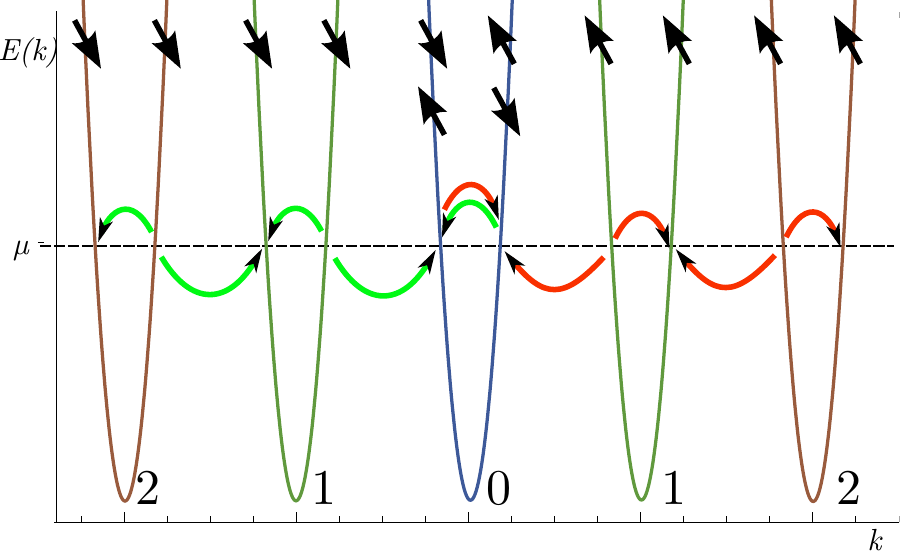}
\caption{\label{fig:parabolas}Dispersion relations $E_j(k)$ for the different wires. Due to the Rashba term, the energies 
of different eigenstates are displaced to opposite sides around zero momentum. The ever increasing Rashba terms induce a 
different momentum shift for each wire. The arrows in the top of each parabola indicate the corresponding spin projection. Around the Fermi level,
tunneling processes are allowed (see main text) and they are represented by red (green) arrows for the $+$ ($-$) spin
states.}
\end{figure}

Using $\psi_{j,a}$ as a basis, we proceed to linearize the Hamiltonian (\ref{Ham_Rashba}) around the Fermi energy.
The linearized Hamiltonian is

\begin{eqnarray}\nonumber\label{Ham_Rashba_lin}
 \mathcal{H}_0=v_F^0\sum_{j,s}\int dx\Big\{ \psi^{\dagger R}_{j,s}(-i\partial_x-k_{F,s}^R(j)) \psi^R_{j,s}\\
  -\psi^{\dagger L}_{j,s}(-i\partial_x-k_{F,s}^L(j)) \psi^L_{j,s}\Big\}.
\end{eqnarray}

General density-density interactions between wires are given by
\begin{equation}
 \mathcal{H}_{{\rm FS}}=\frac{v^0_F}{2\pi}\sum_{jk,\eta\eta',ss'}\int dx \rho_{j,s}^{\eta}(x)V_{\eta\eta',ss'}^{jk}\rho_{j',s'}^{\eta'}(x).
\end{equation}

\noindent where $\rho_{j,s}^{\eta}(x)= \psi^{\dagger \eta}_{j,s}\psi^{\eta}_{j,s}$ is the density of electrons 
in the wire $j$, belonging to the layer $s$, moving in the direction $\eta=(R,L)$. Using the
standard bosonization rules described in the appendix (\ref{app:Bos_densities}) the bosonized Hamiltonian reads
\begin{equation}\label{Ham_micro}
 \mathcal{H}=\mathcal{H}_0+\mathcal{H}_{FS}=\sum_{i,j}\int dx (\partial_x \vec{\phi}^i)^{T} M_{ij}(\partial_x \vec{\phi}^j),
\end{equation}

\noindent where $\vec{\phi}_j$ is the vector of bosonic fields at the position $j$, $\vec{\phi}^j=(\varphi^j_{+},\theta^j_{+},\varphi^j_{-},\theta^j_{-})^T$ that are associated with 
fermions from the upper ($\psi^R_{j,+},\psi^L_{j,+}$) and the lower ($\psi^R_{j,-},\psi^L_{j,-}$) layers.

At this stage, the $U(1)\times U(1)$ symmetry, which appeared after the explicit break of $SU(2)$ by the 
spin orbit term, becomes manifest under bosonization as the symmetry under a constant shift of the phase fields $\varphi(x)\rightarrow \varphi(x)+\beta$.

The most general quadratic Hamiltonian with nearest wires interactions  has the following structure
\begin{equation}\label{model_micro}
M_{ij}=M_0\delta_{ij}+M_1\delta_{i,j+1}+M_1^\dagger\delta_{i+1,j}.
\end{equation}

In order to represent  the  Hamiltonian in terms of local link variables, cf. Eq. (\ref{link_fields}),  
we choose  the forward scattering matrix of the following form
\begin{eqnarray}\label{matrices}
 M_0&=&U^\dagger M U+V^\dagger MV,\\
 M_1&=&V^\dagger M U,
\end{eqnarray}
\noindent with $U$ and $V$ given by

 \begin{eqnarray}\nonumber
 U=\left(
\begin{array}{cccc}
 1 & m & 0 & 0 \\
 1 & m & 0 & 0 \\
 0 & 0 & 1 & -m \\
 0 & 0 & 1 & -m
\end{array}
\right), \quad
V=\left(
\begin{array}{cccc}
 1 & -m & 0 & 0 \\
 -1 & m & 0 & 0 \\
 0 & 0 & 1 & m \\
 0 & 0 & -1 & -m
\end{array}
\right).
\end{eqnarray}

For this special forward scattering matrix (\ref{model_micro}) the Hamiltonian 
\begin{equation}
 \mathcal{H}=\sum_{\ell}\int dx (\partial_x \boldsymbol{\bar{\phi}}_\ell)^{T} M(\partial_x \boldsymbol{\bar{\phi}}_\ell),
\end{equation}
is local in terms of the new fields $\boldsymbol{\bar{\phi}}_\ell=(\bar{\varphi}_{\ell,+},\bar{\theta}_{\ell,+},\bar{\varphi}_{\ell,-},\bar{\theta}_{\ell,-})^T$;
\begin{eqnarray}\label{link_fields_TI}
 2\bar{\varphi}_{j+\frac{1}{2},s}&\equiv&\varphi^j_s+\varphi^{j+1}_s+s|m|(\theta_s^j-\theta_s^{j+1}),\\
 2\bar{\theta}_{j+\frac{1}{2},s}&\equiv&\varphi^j_s-\varphi^{j+1}_s+s|m|(\theta_s^j+\theta_s^{j+1}).
\end{eqnarray}
Following Ref. \onlinecite{Teo2014} we name these bosonic degrees of freedom link fields (note that now they
have an extra layer index $s$). From the definition of the link fields and the commutation relations of the bosonic fields 
$\theta$ and $\varphi$, defined in (\ref{CR}), we find the link fields' commutation relations

\begin{equation}
 [\bar{\varphi}_{\ell,s}(x),\bar{\theta}_{\ell',s'}(x')]=i\pi s|m|{\rm sgn}(x-x')\delta_{s s'}\delta_{\ell \ell'}.
\end{equation}

The quasiparticle charge density and current on the link $\ell$ and layer $s$ are defined accordingly

\begin{eqnarray}\label{quasip_current}
 j^0_{Q,\ell,s}&=&\rho_{Q,\ell,s}=\frac{s\partial_x\bar{\theta}_{\ell,s}}{\pi |m|}\quad\mbox{and}\\
 j^1_{Q,\ell,s}&=&j^{x}_{Q,\ell,s}=\frac{-s\partial_\tau\bar{\theta}_{\ell,s}}{\pi |m|}.
\end{eqnarray}

Note that the link fields defined above transform under time reversal as 

\begin{equation}\label{TRlinks}
 \mathcal{T}\bar{\varphi}_{\ell,s}\mathcal{T}^{-1}=-\bar{\varphi}_{\ell,-s}+\frac{(1-s)\pi}{2},\quad \mathcal{T}\bar{\theta}_{\ell,s}\mathcal{T}^{-1}=-\bar{\theta}_{\ell,-s}
\end{equation}

The $U(1)\times U(1)$ symmetry of the original fermions becomes a symmetry under constant shifts in the links
phase field $\bar{\varphi}_{\ell,s}$. This is no longer the case when the symmetry is gauged. This leads to interesting
consequences for the low energy theory, described in the next sections.

Using (\ref{TRlinks}) one can show that  the most general form of the forward scattering matrix $M$, consistent with time 
reversal symmetry, is given by 
\begin{equation}
 M=\left(
\begin{array}{cccc}
 \alpha_{\varphi\varphi} &  \alpha_{\varphi\theta} &  \alpha_{\varphi\bar{\varphi}} &  \alpha_{\varphi\bar{\theta}} \\
 \alpha_{\varphi\theta} &  \alpha_{\theta\theta} &  \alpha_{\varphi\bar{\theta}} &  \alpha_{\theta\bar{\theta}} \\
 \alpha_{\varphi\bar{\varphi}} & \alpha_{\varphi\bar{\theta}} &  \alpha_{\varphi\varphi} &  \alpha_{\varphi\theta} \\
 \alpha_{\varphi\bar{\theta}} &  \alpha_{\theta\bar{\theta}} &  \alpha_{\varphi\theta} &  \alpha_{\theta\theta}
\end{array}
\right)\,,
\end{equation}

\noindent being parametrized by six independent variables ($\alpha_{\varphi\varphi}$, $\alpha_{\varphi\theta}$, $\alpha_{\varphi\bar{\varphi}}$,
$\alpha_{\varphi\bar{\theta}}$, $\alpha_{\theta\bar{\theta}}$, $\alpha_{\theta\theta}$).
In terms of the link fields the  Euclidean action is given by the sum of over all the link fields
\begin{equation}
 S_0=\sum_\ell\int dxd\tau i(\partial_\tau \boldsymbol{\bar{\phi}}_\ell)^{T} K(\partial_x \boldsymbol{\bar{\phi}}_\ell)-\mathcal{H}\,.
\end{equation}
\noindent Here $K$ is the matrix
\begin{equation}
 K= \frac{1}{2\pi |m|}\left(
\begin{array}{cc}
\sigma_x & 0 \\
0 & -\sigma_x 
\end{array}
\right)\,.
\end{equation}

After integrating out the bosonic phase fields $\bar{\varphi}_{\ell,s}$, we are left with the action for the fields 
$\bar{\theta}_{\ell,s}$ describing density fluctuation on the links. This action can be cast in a familiar form by 
using the charge and spin fields
\begin{eqnarray}\label{ch_spin}
 \Theta_{\ell,+}=\frac{\bar{\theta}_{\ell,+}+\bar{\theta}_{\ell,-}}{\sqrt{2}},\quad  \Theta_{\ell,-}=\frac{\bar{\theta}_{\ell,+}-\bar{\theta}_{\ell,-}}{\sqrt{2}}.
\end{eqnarray}
In terms of $\Theta_{\ell,\pm}$ the action becomes (summation  over $\alpha=+,-$ and links $\ell$ is implied)
\begin{eqnarray}\label{ac_ch_spin}\nonumber
S_0=\frac{1}{2\pi}\int dxd\tau\Big[\frac{u_\alpha}{K_\alpha}(\partial_x{\Theta}_{\ell,\alpha})^2+\frac{1}{K_\alpha u_\alpha}(\partial_\tau{\Theta}_{\ell,\alpha})^2\\
+ic((\partial_x{\Theta}_{\ell,+})(\partial_\tau{\Theta}_{\ell,-})+(\partial_\tau{\Theta}_{\ell,+})(\partial_x{\Theta}_{\ell,-}))\Big],
\end{eqnarray}
\noindent with Luttinger liquid parameters $K_\pm$ given by

\begin{equation}
\frac{1}{K_\pm}=m\sqrt{\frac{\alpha_{\theta\theta}\pm\alpha_{\theta\bar{\theta}}}{\alpha_{\varphi\varphi}\mp\alpha_{\varphi\bar{\varphi}}}-\frac{(\alpha_{\varphi\theta}\pm\alpha_{\varphi\bar{\theta}})^2}{\alpha_{\varphi\varphi}^2-\alpha_{\varphi\bar{\varphi}}^2}}.
\end{equation}
\noindent The velocities $u_\pm$ in terms of the interaction parameters become
\begin{equation}\nonumber
u_\pm=2\pi m\sqrt{(\alpha_{\varphi\varphi}\mp\alpha_{\varphi\bar{\varphi}})\left[\alpha_{\theta\theta}\pm\alpha_{\theta\bar{\theta}}-\frac{(\alpha_{\varphi\theta}\pm\alpha_{\varphi\bar{\theta}})^2}{\alpha_{\varphi\varphi}\pm\alpha_{\varphi\bar{\varphi}}}\right]},
\end{equation}

\noindent while the parameter $c$ (which explicitly breaks parity) reads

\begin{equation}
 c=\frac{\left(\alpha _{\varphi \theta } \alpha _{\varphi \varphi }-\alpha _{\varphi  \bar{\varphi}}
   \alpha _{\varphi  \bar{\theta}}\right)}{2\pi|m|(\alpha _{\varphi \varphi }^2-\alpha _{\varphi  \bar{\varphi}}^2)}.
\end{equation}

The action (\ref{ac_ch_spin}) is self dual under $\rho\leftrightarrow\sigma$ and 
resembles the action that appears in the context of spin ladders \cite{GogolinBook2004}.

\subsubsection{Tunneling operators}
So far we did not allow for electron tunneling between different wires.
To account for such  processes we consider  the multi-particle hopping   operators
\begin{eqnarray}
 {\mathcal{O}}_{j,a}^{\{s_{p,a}^R,s_{p,a}^L\}}=\prod_p(\psi_{j+p,a}^{R}(x))^{s^R_{p,a}}(\psi_{j+p,a}^{L}(x))^{s^L_{p,a}}.
\end{eqnarray}
Here we use the convention $\psi^{-1}=\psi^\dagger$. The possible choices of integers $s^R$ and $s^L$ are restricted by
charge conservation $\sum_p({s^R_{p,a}}+{s^L_{p,a}})=0$, and momentum conservation
\begin{flalign}
 \sum_ps_{p,a}^Rk_{F,a}^R(j+p)+s_{p,a}^Lk_{F,a}^L(j+p)=0,
\end{flalign}

\noindent replacing the value of the Fermi momentum (\ref{Fermi_mom}) we have

\begin{flalign}\label{mom_cons}
 k^0\sum_p(s_{p,a}^R-s_{p,a}^L)-a\sum_p\Lambda_{j+p}(s_{p,a}^R+s_{p,a}^L)=0.
\end{flalign}

The  solution of  this equation exists  for all values of $j$  only if  $\Lambda_j$ is  linear in $j$. 
This restricts us to  the confining potentials given by Eq. (\ref{confining}).

Setting 
\begin{equation}
\label{special_spin_orbit}
\Lambda_j=j\Lambda
\end{equation}
 we  rewrite Eq.(\ref{mom_cons}) as 
\begin{equation}\label{condition}
\frac{ak^0}{\Lambda}=\frac{\sum_pp(s_{p,a}^R+s_{p,a}^L)}{\sum_p(s_{p,a}^R-s_{p,a}^L)},
\end{equation}

\noindent where we have used that $a=(+,-)$. As $s^R,s^L$ are integers, equation (\ref{condition}) has solutions 
only when $\frac{ak^0}{\Lambda}$ is a rational number. The simplest FTI phase corresponds to 

\begin{equation}
\label{rational}
\frac{ak^0}{\Lambda}=\frac{a}{2m}.
\end{equation}
\noindent This is similar to the condition (\ref{filling_frac}) in the construction of Laughlin states.
In that case, if the filling fraction matches $1/m$, the tunneling operators 
(\ref{tunneling_Laughlin}) conserve the momentum and can be relevant. In analogy with the single layer 
scenario,  eq. (\ref{condition}) corresponds to the condition of an {\it effective} filling fraction $\nu_a=2ak^0/\Lambda$
in the layer $a$.

Any operator  ${\mathcal{O}}_{j,a}^{\{s_{p,a}^R,s_{p,a}^L\}}$, with the set of parameters $\{s_{p,a}^R,s_{p,a}^L\}$, 
satisfying Eq.(\ref{condition}) describes a legitimate multi-particle hopping.  
The most relevant operator in the RG sense is given by 
\begin{eqnarray}
 s_{1,a}^R=-\frac{am-1}{2}\quad s_{1,a}^L=\frac{am+1}{2}\\
  s_{0,a}^R=-\frac{am+1}{2}\quad s_{0,a}^L=\frac{am-1}{2}\\
  s_{p,a}^{R/L}=0 \quad \mbox{for $p\neq 0,1$} 
\end{eqnarray}

\noindent which are integers when $m=\Lambda/2k^0$ is an odd integer.
Upon bosonization, ${\mathcal{O}}_{j,a}^{\{s_{p,a}^R,s_{p,a}^L\}}$ in this case becomes 
\begin{equation}\label{tunneling}
\mathcal{O}_{\ell,s}=\cos(2\bar{\theta}_{\ell,s}),
\end{equation}

\noindent where $\bar{\theta}_\ell$ is defined in Eq. (\ref{link_fields_TI}). The corresponding action 
is given by 
\begin{equation}\label{action_mass}
S=S_0+g\sum_\ell\int dxd\tau(\cos(2\bar{\theta}_{\ell,+})+\cos(2\bar{\theta}_{\ell,-}))\,.
\end{equation}
As we see, in the bosonic notation, the inclusion of tunneling operator leads to the sine-Gordon type action.
In the case when the cosine term is a relevant perturbation, a gap opens in excitation spectrum of the sliding LL
in the bulk. To study this question we realize an (RG) analysis of this operator. 

\section{RG analysis, the bulk gap of FTI and edge modes}

\subsection{Relevance of tunneling operators}

The RG equations can be derived for this problem in the standard way \cite{GogolinBook2004}, and one finds
\begin{eqnarray}\label{RG_eqs}
 \frac{dg}{dl}&=&(2-\Delta)g,\\
\frac{d}{dl}\left(\frac{u_\rho}{K_\rho}\right)&=&\frac{d}{dl}\left(\frac{u_\sigma}{K_\sigma}\right)=\frac{\Delta g^2f(x)}{u_\rho},\\\label{RG_eqs2}
\frac{d}{dl}\left(\frac{1}{u_\rho K_\rho}\right)&=&\frac{d}{dl}\left(\frac{1}{u_\sigma K_\sigma}\right)=\frac{\Delta g^2f(x)^3}{u_\rho^3}.
\end{eqnarray}
Here $x \equiv u_\rho/u_\sigma$ is the  ratio of sound velocities in the charge and spin sector. The functions $f(x)$ and
$h(x)$ are
\begin{eqnarray}\label{f_function}
 f(x)&=&\sqrt{h(x,c)+\sqrt{h(x,c)^2-x^2}},\\
 h(x,c)&=&(1+x^2+|c|^2x)/2.
\end{eqnarray}
The scaling dimension of the tunneling operator $\mathcal{O}_{\ell,s}$ is given by
\begin{equation}
 \Delta[\cos(2\bar{\theta}_{\ell,s})]=\frac{K_\rho+K_\sigma}{2\sqrt{1+\frac{|c|^2x}{(1+x)^2}}}.
\end{equation}
There is no renormalization of the parameter $c$ up to order $g^2$.

The set of equations (\ref{RG_eqs}-\ref{RG_eqs2}) can be linearized around the fixed point $(\Delta^0,g^0)=(2,0)$.
The fixed point $\Delta(K_\rho,K_\sigma,x)=2$ defines a region (in parameter space) 

\begin{equation}\label{surface}
\left(\frac{K_\rho+K_\sigma}{4}\right)^2=1+\frac{|c|^2x}{(1+x)^2}, 
\end{equation}

\noindent for the LL parameters and velocities. Expanding around the point $\mathbf{p}^0=(K^0_\rho,K_\sigma^0,x^0)$ 
belonging to the surface defined in Eq. (\ref{surface}),  
 we define $\Delta\equiv 2+\lambda$ with $\lambda\ll1$ and $y\equiv g/u_\rho^*$ (see also appendix (\ref{sec:RG})).
The set of RG equations (valid in the vicinity of the fixed  point), is given by 
\begin{eqnarray}\label{RG_lineareqs}
 \frac{dy}{dl}&=&-\lambda y,\\
 \frac{d\lambda}{dl}&=&-\mathcal{C} y^2,\label{eq:RG_quadratic}
\end{eqnarray}
where the terms quadratic in $\lambda$ were neglected.
\noindent The parameter $\mathcal{C}=\mathcal{C}(K_\rho^0, K_\sigma^0,x^0)$ is an involved function of the point $\mathbf{p}^0=(K^0_\rho,K_\sigma^0,x^0)$  
which belongs to the surface (\ref{surface}).
 
For $\mathcal{C}\leq 0$, the cosine term in Eq. (\ref{tunneling}) is never relevant and the system  goes into the 
weak coupling fixed point. In a generic situation $\mathcal{C}>0$  and the tunneling operators 
become relevant for $\lambda<0$. This corresponds to a Berezinsky-Kosterlitz-Thouless (BKT) transition at $\Delta<2$ (see also (\ref{f_RG})).

\begin{figure}
\includegraphics[width=1\columnwidth]{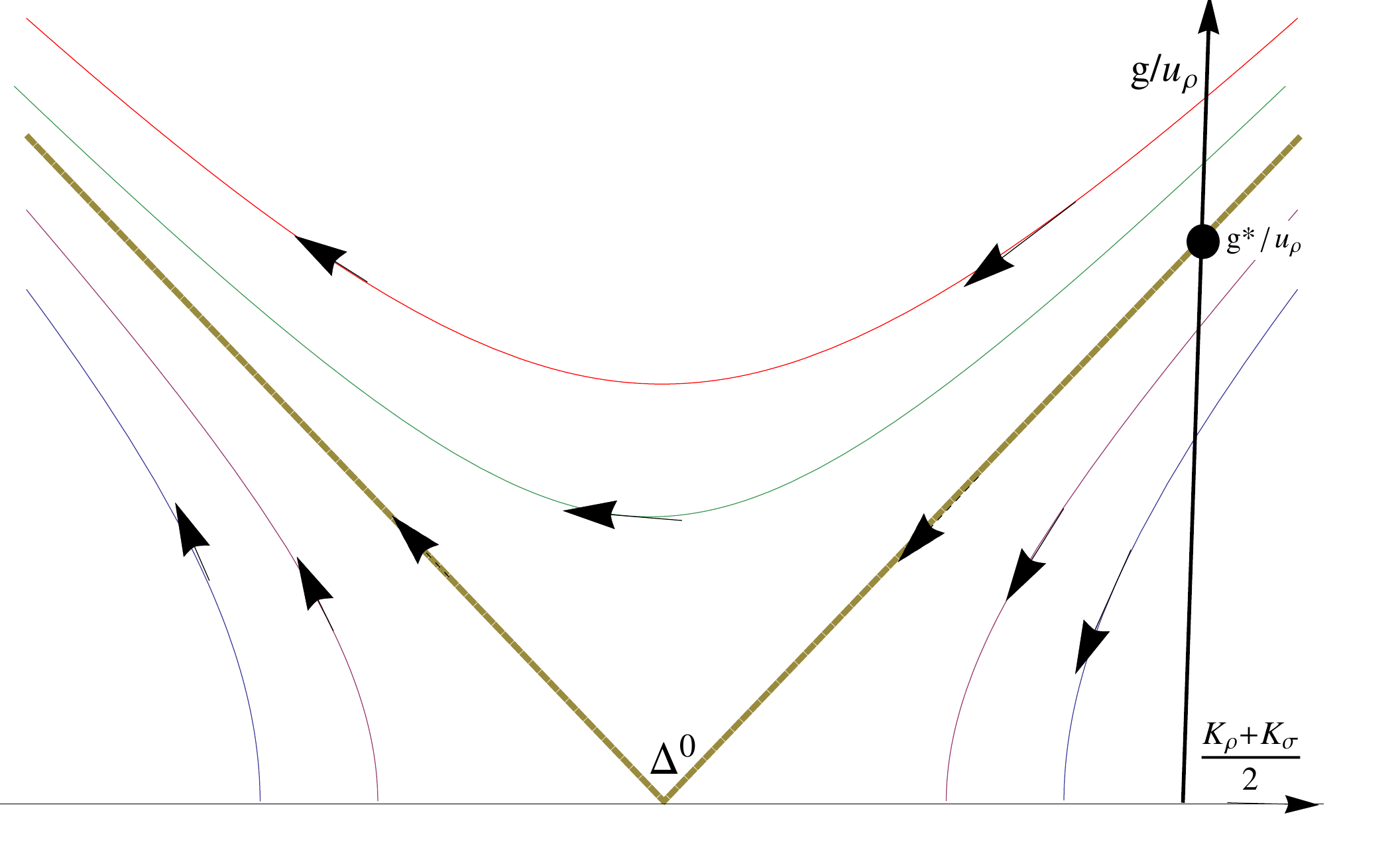}\caption{\label{fig:BKT} Renormalization flows for Eqs. 
(\ref{RG_lineareqs}
and \ref{eq:RG_quadratic}). The different lines indicate the different
initial strengths of $g/u_{\rho}$. The thick dashed line represents
the separatrix of the BKT transition. When the initial $g/u_{\rho}$
is above $g^*/u_{\rho}$, the system flows to the strong coupling
regime (fractional topological phase), while for initial values below $g^*/u_{\rho}$,
the system flows to weak coupling regime (sliding LL). This plot is made for $c=0.1$.}
\end{figure}

As shown in Fig. \ref{fig:BKT}, this implies that the system
flows to strong coupling regimes if
\begin{equation}
\frac{K_\rho+K_\sigma}{2}<2\sqrt{1+\frac{|c|^2x}{(1+x)^2}}.
\end{equation}

\begin{figure}
(a)

\includegraphics[width=1\columnwidth]{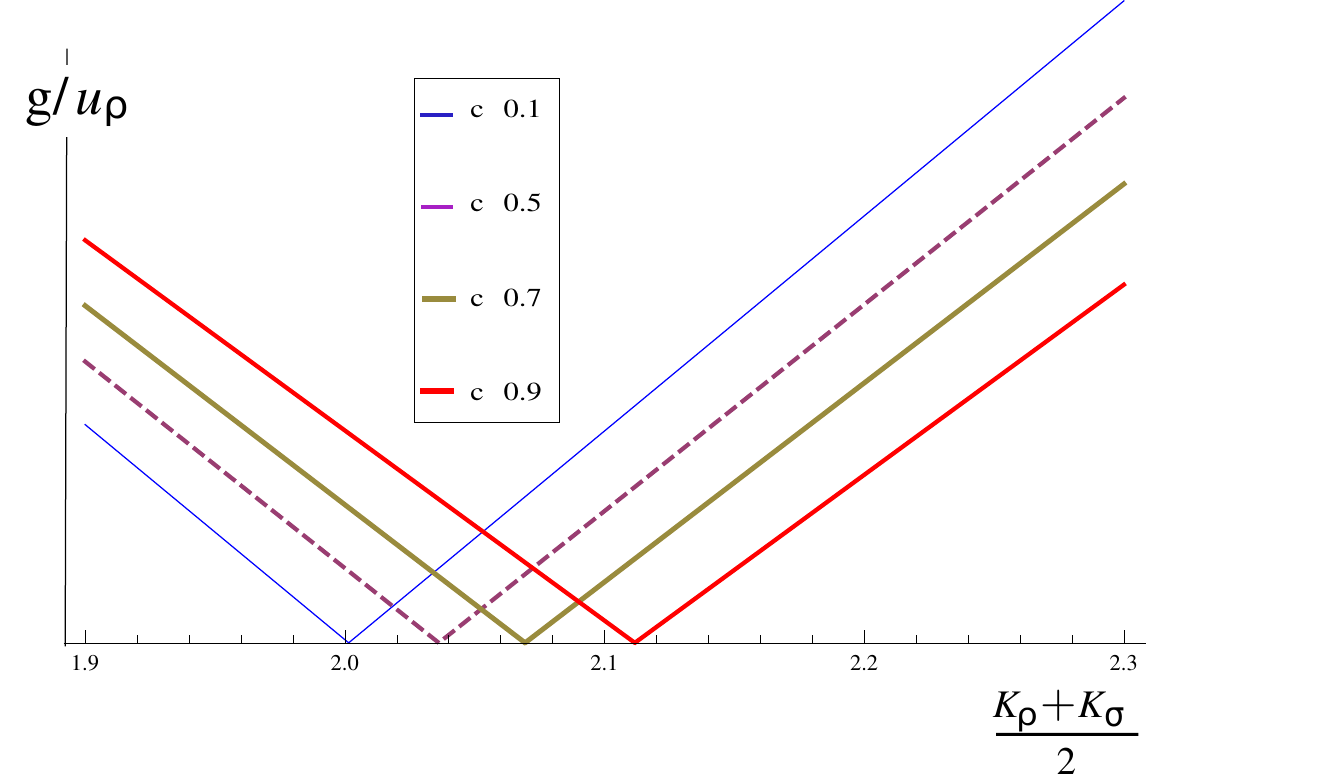} 

(b)

\includegraphics[width=1\columnwidth]{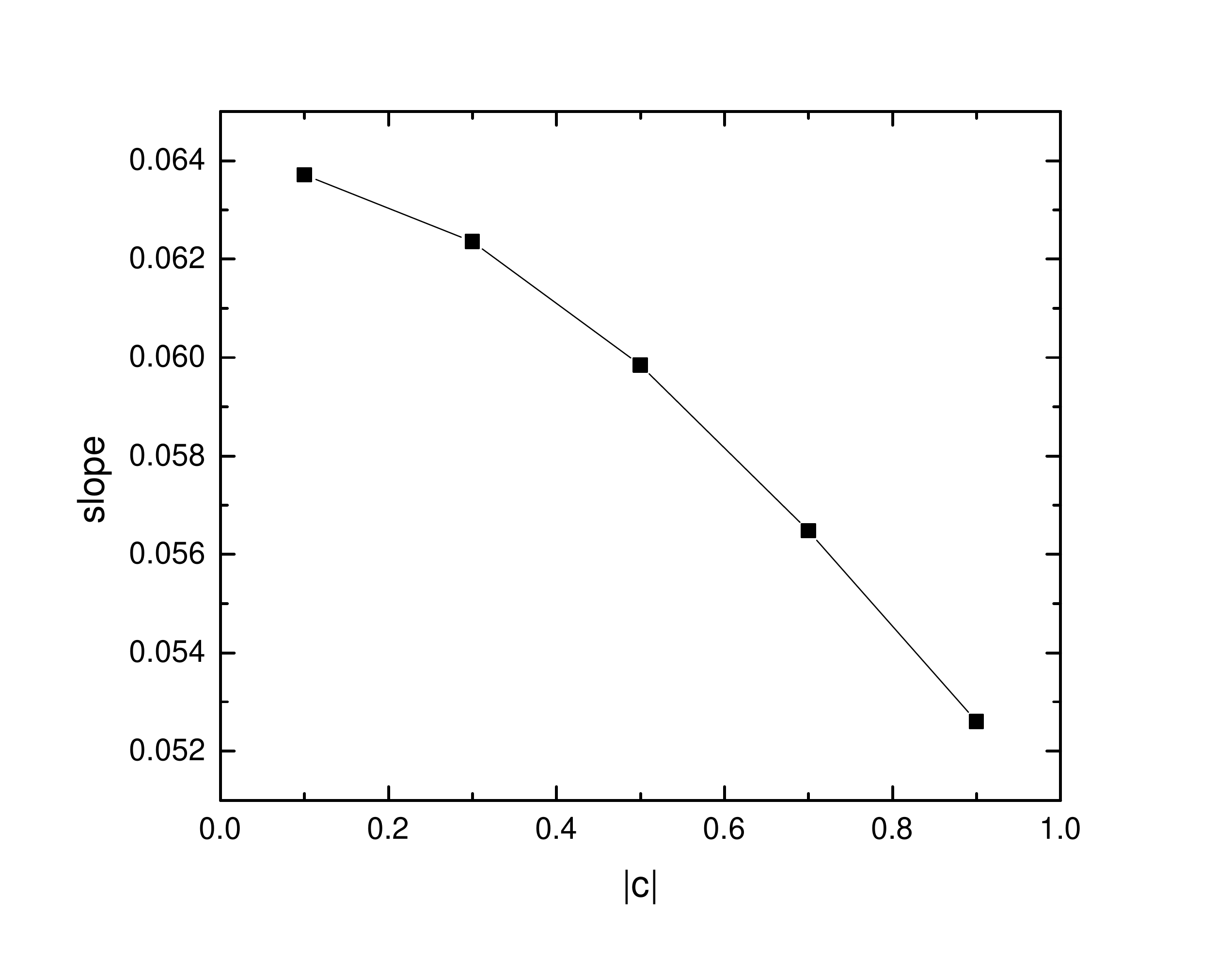}

\caption{\label{fig:separatrix} Various separatrix lines with various interaction
strengths c. (a) For a generic electron interaction (finite $c$) the critical point $\Delta$ shifts
to the right with increasing $c$. (b) The slope of the
separatrix lines decreases as a function of $c$,  that means that the region where the system flows to strong coupling (i.e the topological phase)
is enlarged.}
\end{figure}

For $c=0$, the theory becomes massive when the LL parameters $K_\rho+K_\sigma<4$. For generic interactions 
(preserving time reversal symmetry), $c\neq0$ and the region where the Luttinger parameters flow to strong coupling
is enhanced as shown in Fig  \ref{fig:separatrix}. In this sense, generic time reversal symmetric interactions help 
driving the system into the topological phase.

From now on, we assume that the system is in the massive phase. Our RG eqs. show that the cosine term in Eq. (\ref{tunneling}) 
is relevant and a gap in the spectrum of sliding LL opens (this was just assumed in Ref. \onlinecite{Klinovaja2014b}). 
For periodic boundary conditions, the gap opens everywhere, as the field $\phi_s^N=\varphi_s^N-s|m|\theta_s^N$ in the wire 
$N$ pairs up with the field $\phi_s^0=\varphi_s^0+s|m|\theta_s^0$ in the first wire to form the link fields, all of which 
develop a gap after the cosine term becomes RG relevant (see eq. (\ref{link_fields_TI})).

\subsection{Edge modes}

Experimentally however, a two dimensional finite size sample is modeled by open boundary conditions. In this case 
note that the first and last fields (with different \textquotedblleft layer" components) 

\begin{eqnarray}\label{edgemodes}
 \phi^s_0&\equiv&\varphi_s^0+s|m|\theta_s^0,\\
 \phi^s_N&\equiv&\varphi_s^N-s|m|\theta_s^N,
\end{eqnarray}

\noindent  had no cosine term and therefore remain ungapped.
These fields at the boundary correspond to the gapless edge modes of a TI.
The same conclusion has been reached in Ref. \onlinecite{Klinovaja2014b}. 
At each edge we have two counter propagating edge modes, labeled by $s=(+,-)$. 
The dynamics for these edge modes is described by the Hamiltonian

\begin{equation}\label{edge_ham}
 H_{\rm edge}=\frac{|m|v_F}{4\pi}\sum_{s=+,-}\int dx(\partial_x\phi^s_0)^2 +(\partial_x\phi^s_N)^2.
\end{equation}

These helical edge modes  are  the hallmark of the topologically nontrivial phase. We will see in the next section that the phase that we encountered here is described in the long wavelength limit by a Chern-Simons theory, which makes the topological  properties more evident.

Different boundary conditions can be implemented using different arrangements of wires. In order to reproduce chiral states
appearing in the edge of a disk, an array of concentric wires can be used, as discussed in the appendix \ref{app:CW}.

Few comments are in order here.
First, we note  that our analysis  is valid  only for  special filling fractions,  formally described by Eq. (\ref{condition}). 
Moreover it is limited  by  the special type of spin-orbit interaction in Eq. (\ref{special_spin_orbit}), and the confining 
potential in Eq. (\ref{confining}).
As we are about to see, in this case the system has topological order. 
Although we do not address the stability of this point, we expect that static disorder stabilizes  
the topological phase, making it robust against small deviations on the filling factor and generic spin-orbit interactions.

Our analysis is limited to the Laughlin states with an {\it effective} filling factor $\nu=\pm1/m$ , $m$ being an odd integer.
A heedful reader could ask, what does prohibit us from repeating this procedure for even values of  $m$. 
While formal steps of construction remain valid, the result  is not the true ground state. 
As we know from the context of fractional quantum Hall effect \cite{WenBook2004}  
the true ground state for these filling  is highly sensitive to the fine interaction details,
and thus requires special consideration,  that  is  beyond the current work.

We now show how our approach can be connected with  a more conventional Chern-Simons theory.

\section{Low energy description and Chern-Simons term}
Until now, we have considered the dynamics of fermionic excitations in the system. We will now discuss the 
electromagnetic response of the system and the effective low energy description of the FTI state.
As expected for the TI case, the resulting theory indeed has a topologically non-trivial Chern-Simons term. 

To find out the electromagnetic response, one needs to compute the effective action $I$ as follows, 
\begin{equation}
 \exp(-I[A])=\int\mathcal{D}{\psi}\mathcal{D}\bar{\psi}e^{-S[\psi,\bar{\psi},A]}.
\end{equation}

\noindent Here $S$ is the fermionic action in the presence of an external gauge field $A$.
The coupling of the gauge field to the fermionic action in its bosonic form in Eq. (\ref{action_mass})
should be gauge invariant. We proceed to investigate the gauge invariance of $S$ below.

\begin{subsection}{Gauge invariance}
\end{subsection}

Let's recall the gauge transformation laws for the gauge field $A_\mu$ and original fermions $\psi_j(x,\tau)$.
Due to charge conservation, the phase of electron operators can be redefined $\psi'(x,\tau)=e^{i\alpha}\psi(x,\tau)$, without 
changing the action. This $U(1)$ freedom can be {\it gauged} by making the phase $\alpha$ (initially a constant) a function
of space-time. Of course, this change alone does not preserve the fermionic action. To make the action invariant again 
it is necessary to include a gauge field with transformation laws that cancel the extra contribution
from the space-time dependent phase. We have then the following gauge transformations
\begin{eqnarray}\label{GTfermion}
 [\psi_j(x,\tau)]^G&=&e^{i\alpha_j(x,\tau)}\psi_j(x,\tau),\\\nonumber
 [A_{\mu,j}(x,\tau)]^G&=&A_{\mu,j}(x,\tau)-\partial_\mu\alpha_j(x,\tau).
\end{eqnarray}
It is easy to check that these transformations leave invariant the free electron action
coupled to an external gauge field
\begin{eqnarray}
 S_{\rm free}[\psi,A]=\int d^2x\bar{\psi}(i\partial_\mu\gamma^\mu-A_\mu\gamma^\mu)\psi,
\end{eqnarray}

\noindent i.e. $S_{\rm free}[\psi^G,A^G]=S_{\rm free}[\psi,A]$, (here $\bar{\psi}=\psi^\dagger\gamma^0$ and $\gamma^{0,1}$ 
are the Dirac matrices defined in (\ref{ap:TR})).

Bosonizing the electron (see (\ref{app:bosonization})), we can read off the gauge transformation rules for the phase 
$\varphi$ and density field $\theta$

\begin{eqnarray}
 [\varphi^j_s(x,\tau)]^G&=&\varphi^j_s(x,\tau)+\alpha_j(x,\tau),\\\nonumber
 [\theta^j_s(x,\tau)]^G&=&\theta^j_s(x,\tau).
\end{eqnarray}

The link fields in Eq. (\ref{link_fields_TI}), that appear as the natural basis after the inclusion of the tunneling operators (\ref{tunneling}),
change non trivially under a gauge transformations

\begin{eqnarray}\nonumber
 [\bar{\varphi}_{j+\frac{1}{2},s}(x,\tau)]^G&=&\bar{\varphi}_{j+\frac{1}{2},s}(x,\tau)+\frac{\alpha_j(x,\tau)+\alpha_{j+1}(x,\tau)}{2},\\\nonumber
 [\bar{\theta}_{j+\frac{1}{2},s}(x,\tau)]^G&=&\bar{\theta}_{j+\frac{1}{2},s}(x,\tau)+\frac{\alpha_j(x,\tau)-\alpha_{j+1}(x,\tau)}{2}.
\end{eqnarray}

\noindent Note in particular that $\bar{\theta}_{j+\frac{1}{2},s}$ transforms as a gauge field with $\alpha_j-\alpha_{j+1}$
the discretized version of the derivative along $y$.

The gauge field parallel to the links, along with the temporal component (scalar potential) couples to the action
via the current and density operators. After bosonization the part of the action that couples the gauge field to the 
quasiparticles is explicitly

\begin{equation}\label{electric_f}
 S_A=\sum_{\ell,s,}\int d^2xj_{\ell,s}^\mu A_{\mu}=\sum_{\ell,s}s\int d^2x \frac{\partial_\nu\bar{\theta}_{\ell,s}}{|m|\pi}\epsilon^{\mu\nu} A_{\parallel\mu},
\end{equation}

\noindent where $j_{\ell,s}^\mu $ is the quasiparticle density ($\mu=0$) and current ($\mu=1$), defined in 
(\ref{quasip_current}); $\epsilon^{\mu\nu}$ is the two dimensional Levi Civita antisymmetric tensor, with $\epsilon^{01}=1$
and summation over repeated indices is assumed. The gauge field along the wires is $A_\parallel=(A_0(\tau,x),A_1(\tau,x))$ 
($\mu=(0,1)=(\tau,x)$ in (\ref{electric_f})).

The last term in Eq. (\ref{electric_f}) can be integrated by parts, and ensures that the action is gauge invariant with respect 
to the temporal and the $x$-component of the gauge 
field. The gauge field $A^{\perp}=A_2$ which is perpendicular to the wires induces an Aharonov-Bohm phase in 
the inter-wire tunneling operator $\mathcal{O}_\ell$ defined in Eq. (\ref{tunneling}).
Indeed, the tunneling operator for the $s=+$ (upper layer) fields in terms of the original fermions is given by
\begin{equation}
 \mathcal{O}_{\ell,+}=(\psi_{j+1,+}^{L\dagger})^{\frac{m+1}{2}}(\psi_{j+1,+}^{R})^{\frac{m-1}{2}}(\psi_{j,+}^{L\dagger})^{\frac{m-1}{2}}(\psi_{j,+}^{R})^{\frac{m+1}{2}}\,,
\end{equation}
and similarly for the lower layer (with $m$ replaced by $-m$). Under the gauge transformations (\ref{GTfermion})
the tunneling term $\mathcal{O}_\ell$ changes as

\begin{equation}\label{GTO}
 [\mathcal{O}_{j+\frac{1}{2},s}]^G=\mathcal{O}_{j+\frac{1}{2},s} e^{i(\alpha_j-\alpha_{j+1})}.
\end{equation}

In order to maintain gauge invariance, the operator $\mathcal{O}_{\ell,s}$ should be modified, introducing a
{\it Wilson line} $W_{j,j+1}$

\begin{equation}
 W_{j,j+1}=\exp\left(-i\int_{j}^{j+1}A_{y,j} dy\right),
\end{equation}

\noindent such that the tunneling operator $\mathcal{O}_{\ell,s}$ becomes

\begin{equation}\label{M_tunn}
 \mathcal{O}_{j+\frac{1}{2},s}\rightarrow \mathcal{O}_{j+\frac{1}{2},s}W_{j,j+1}.
\end{equation}

It is clear that this combination is invariant under a gauge transformation, as the
phase acquired by $\mathcal{O}$ is exactly canceled by the transformation of $W$

\begin{eqnarray}
 [W_{j,j+1}]^G&=&\exp\left(-i\int_{j}^{j+1}[A_{y,j}]^G dy\right),\\\nonumber
 &=&\exp\left(-i\int_{j}^{j+1}(A_{y,j}-\partial_y\alpha_j)dy\right),\\\nonumber
 &=&W_{j,j+1}e^{-i(\alpha_j-\alpha_{j+1})}.
\end{eqnarray}

The bosonized action (including the modified tunneling term (\ref{M_tunn})) becomes

\begin{eqnarray}\nonumber\label{cos_action}
 S[A]&=&S_0+\sum_{\ell,s}\int d^2x g\cos(2(\bar{\theta}_{\ell,s}+A_\ell^{\perp}d/2))\\
 &-&\sum_{\ell,s}\int d^2x\frac{\bar{\theta}_{\ell,s}}{|m|\pi}\epsilon^{\mu\nu}\partial_\mu A_{\parallel\nu}.
\end{eqnarray}

\noindent This action is gauge invariant (here $A^\perp_\ell=A_{y,j}$ and we have used that $\int A_y dy=A_yd$, where $d$ is the
distance between wires which is assumed smaller than any other length scale).

When the cosine term becomes relevant, the field $\bar{\theta}_\ell$ is pinned to the minimum of the potential, i.e
$\bar{\theta}_\ell=-A_\ell^\perp d/2$. The action acquires a term proportional to the chiral anomaly in each
effective layer $(+,-)$, and this anomaly dominates the dynamics of the system after the massive modes ($\theta$ fields)
have been integrated out \cite{Mulligan2011}. 
In this case the effective action for the upper layer becomes (using $\Delta y=d$)

\begin{equation}\label{CS_discrete}
I^+[A]=\frac{1}{2\pi|m|}\sum_\ell\Delta y\int d^2x\epsilon^{\mu\nu}A^{\perp}_\ell\partial_\mu A_{\parallel\nu}+\dots.
\end{equation}

\noindent where $\dots$ denote terms containing higher derivatives, that do not contribute to the small momentum
behavior. Here we recognize the discretized version of (see  Eq. (\ref{app:CS}) for more details)
\begin{equation}\label{ch_anomalies}
I^+[A]=\frac{1}{4\pi|m|}\int d^3x\epsilon^{\mu\nu\rho}A_\mu\partial_\nu A_{\rho}\,, 
\end{equation}
that is the Chern-Simons theory for the fractional quantum Hall effect with filling fraction $\nu=1/m$.
The effective action $I^+$ accounts for the low energy dynamics of the system, 
describing the electromagnetic response  with respect to the gauge field  $A_\mu$ in the upper layer
(similar construction can be done for the lower layer separately). 
Taking the functional derivative respect to the gauge field $A_\mu$, one expectedly reproduces the Hall conductance 
\begin{equation}\label{eq_mov}
\frac{\delta I^+[A]}{\delta A_\mu}\equiv J_+^\mu=\frac{1}{2\pi|m|}\epsilon^{\mu\nu\rho}\partial_\nu A_\rho,
\end{equation}
\noindent where $J_+^\mu$ denotes the electron current in the upper layer. For
the lower layer we have similarly

\begin{equation}
I^-[A]=\frac{-1}{4\pi|m|}\int d^3x\epsilon^{\mu\nu\rho}A_\mu\partial_\nu A_{\rho}\,.
\end{equation}
To satisfy (identically)
the conservation equation $\partial_\mu J^\mu=0$,
it is customary \cite{WenBook2004} to define the bosonic $b^+_\mu$ field by $J^\mu=\frac{1}{2\pi}\epsilon^{\mu\nu\rho}\partial_\nu b_\rho^+$
(note that these auxiliary bosonic fields $b_\mu$ are essentially the full gauge invariant version of 
$\epsilon^{\mu\nu}\partial_\nu\bar{\theta}_\ell$).
The field theory in terms of $b^+_\mu$ that reproduces (\ref{eq_mov}) is then \cite{Lopez1991}

\begin{align}\label{eff_action}\nonumber
 \mathcal{L}_{{\rm b}}[b^{\pm},A]=-\frac{|m|}{4\pi}\epsilon^{\mu\nu\rho}(b^{+}_\mu\partial_\nu b^{+}_\rho-b^{-}_\mu\partial_\nu b^{-}_\rho)\\
 -\frac{1}{2\pi}\epsilon^{\mu\nu\rho}A_\mu(\partial_{\nu}b_\rho^{+} +\partial_{\nu}b_\rho^{-}).
\end{align}

\noindent where we have included the contribution from both layers. Note that the theory (\ref{eff_action}) can be 
cast in the form of the BF theory \cite{Cho2011,Chan2013}, by defining the fields $a_\mu=b_\mu^++b_\mu^-$ and $b_\mu=b_\mu^+-b_\mu^-$.
This double Abelian Chern-Simons theory $U(1)_m\times \overline{U(1)}_m$ accounts for the ground state
degeneracy $m^g$ of the system if placed on a genus $g$ surface. This theory also describes the exchange statistics of 
quasiparticle excitations \cite{WenBook2004,FradkinBook2013,Freedman2004}.

\section{Summary and Outlook}\label{conclusions}
 
Starting from an array of parallel LLs with Rashba type spin-orbit interaction, 
we have constructed a FTI.
The magnitude of spin-orbit interaction depends on the position of the wire and mimics
the  magnetic field in the quantum Hall effect. The  interaction between electrons is restricted to 
nearest and the next nearest wire, but includes all terms  consistent  with $\mathcal{T}$ symmetry.
We have derived and analyzed the RG equations for the multi-particle tunneling operators and showed 
that there exist a window  in the parameter space where  these operators are relevant, and FTI is a stable fixed point.
Due to the large number of possible interaction constants the precise position of 
this window can be determined only numerically.
We have analyzed the stability of the FTI fixed point in relation to a different interaction constants.
Remarkably, the stability is enhanced  by inclusion the repulsive interaction between electron with opposite spin.


To establish the connection with other methods,  we  have derived an  effective low energy theory for our model.
The resulting fixed point of our construction is captured by a double Chern-Simons theory with correct counting of topological ground state degeneracy and exchange of quasi-particle excitations, and agrees with one  
suggested in Ref.\onlinecite{Levin2009} within a more restrictive model.

Before ending the paper, we list some of the questions that remain  open:

a) The stability of the FTI against perturbations that break TR symmetry.
Though a strong perturbation of this kind certainly destroys the FTI state, 
the topological  protection may  hold against  "weak" perturbations.
For the  magnetic impurities located  solely on the edge, the BKT transition between TI and trivial insulator 
was found\cite{Beri2012}. The case  of  magnetic impurities added in the bulk remains to be studied.


b)  Within the sliding LL approach the tunneling operators are chosen for the  
states exactly at fractional (or integer) fillings. In the presence of TR  preserving disorder  
one expects the state to form  a plateau.   
The  emergence of plateaus yet remains to be shown within sliding LL approach.
While for non-interacting case the formation of plateaus and the transition between different filling is described 
within  $\sigma$-model approach\cite{Pruisken1988},  
the current formalism provides a natural framework in the presence of interactions.

 
\medskip

\textbf{Acknowledgments} 
We thank  S.T. Carr, E. Berg,  F. D. M. Haldane, A. D. Mirlin,  I.V. Protopopov,  and A. Stern for the useful discussions. 
The authors are specially indebted to Eran Sela for very illuminating discussions.
This work was  supported by grant GIF 1167-165.14/2011, ISF and DFG grant RO 2247/8-1.


%

\appendix
 
\widetext
 
\section{}

\subsection{Bosonization}\label{app:bosonization}

Under bosonization, a collection of the fermionic operators labeled by index $j$, of chirality $\eta=(R,L)=(+,-)$ and
spin $s=(\uparrow,\downarrow)=(+,-)$ around the Fermi point $k_{F,j}^{\eta,s}$ become \cite{Coleman1975,GiamarchiBook2003}
\begin{equation}\label{bosonization}
 \psi^\eta_{j,s}(x,t)=\frac{U^j_{\eta,s}}{\sqrt{2\pi x_c}}e^{i(k_{F,j}^{\eta,s}x+\varphi^j_s(x,t)+\eta\theta^j_s(x,t))}.
\end{equation}

\noindent The bosonic operators $\theta_s^j(x,t)$ and $\varphi_s^j(x,t)$ follow equal time commutation relations

\begin{equation}\label{CR}
 [\varphi^j_s(x,t),\theta^{j'}_{s'}(x',t)]=i\pi{\rm sgn}(x-x')\delta_{ss'}\delta_{jj'},
\end{equation}

\noindent and $U^j_{\eta,s}$ is a Klein factor that ensure anticommutation of fermions. In this notation, chiral fields 
$\phi^{j}_{R,s}$ (right mover) and $\phi^{j}_{L,s}$ (left mover) are 

\begin{equation}
 \phi^{j}_{R,s}=\varphi^j_s+\theta^j_s,\quad\mbox{and}\quad\phi^{j}_{L,s}=\varphi^j_s-\theta^j_s.
\end{equation}

The bosonization of electron densities become

\begin{equation}\label{app:Bos_densities}
 \rho_{j,s}^R=(\psi^R_{j,s})^\dagger\psi^R_{j,s}=\frac{1}{2}(\partial_x\theta^j_{s}+\partial_x\varphi^j_{s}) 
 \quad\mbox{and}\quad \rho_{j,s}^L=(\psi^L_{j,s})^\dagger\psi^L_{j,s}=\frac{1}{2}(\partial_x\theta^j_{s}-\partial_x\varphi^j_{s}).
\end{equation}

\subsection{Set of concentric loops and tunneling operators}\label{app:CW}
In this appendix we show how the wire construction can be done for a disc geometry.
\begin{figure}[tb]
  {\includegraphics[width=0.5\linewidth]{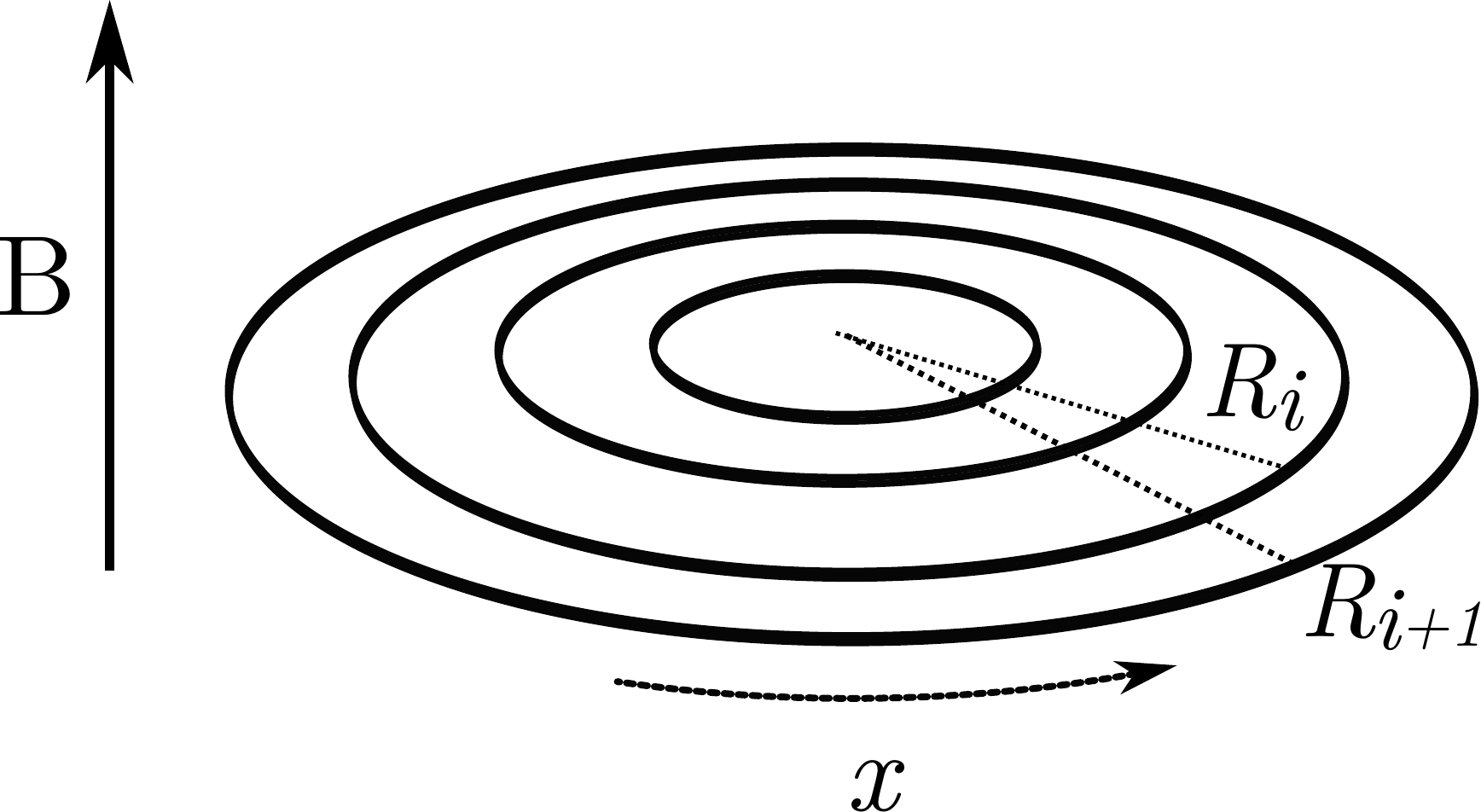}}
	\caption{}\label{fig:cir_wires}
\end{figure}
We choose a set of $N_w$ concentric wires (see Fig.\ref{fig:cir_wires}), with linearly increasing radius
(i.e. radius of circle $j$ being $R_j={j}R_0$).
We classify the states by  integer $n$, the component of angular momentum component in the direction  perpendicular to the 
disc. The geometry of the problem makes it convenient to use the  he symmetric gauge $\vec{A}(r)=\frac{Br}{2} \hat{ \theta}$.
Solving the Schr{\"o}dinger equation for spinless electrons one finds energy levels for the wire $j$ 
\begin{equation}\label{non_int_energy_wires}
 E_n^{(j)}=\frac{\hbar^2}{2M_e R_0^2}\left(\frac{n}{j}-j\frac{eBR_0^2}{2c\hbar}\right)^2.
\end{equation}
It is clear that  this set of wires also   coincide with  the positions of guiding centers  
of the  wave functions for the same gauge, Landau level and angular momentum.
The sliding LL construction can  be perceived either as specially fabricated setup, 
or as a complete basis of highly anisotropic states.

For a given chemical potential $\mu=\frac{\hbar^2(n_F^0)^2}{2M_e R_0^2}$, 
the Fermi (angular) momentum in the wire $j$ is 
$n_{F,j}^\pm=\pm jn_F^0+ j^2\Phi(R_0)$, with $\Phi(R)=\frac{B\pi R^2}{(h/e)c}$ the magnetic flux in units of the
flux quantum. Note that $n_{F,j}$ is well defined just for $\Phi(R_0)\in \mathbb{Z}$.







We are interested in the tunneling terms of the form 

\begin{equation}\label{tun_op}
 \mathcal{O}_{j,j+1}=(\psi^{\dagger}_{L,j+1})^{\frac{m+1}{2}}(\psi_{R,j+1})^{\frac{m-1}{2}}(\psi^{\dagger}_{L,j})^{\frac{m-1}{2}}(\psi_{R,j})^{\frac{m+1}{2}},
\end{equation}

\noindent in terms of bosonic degrees of freedom $\varphi,\Theta$ the tunneling operator (\ref{tun_op}) becomes

\begin{equation}
 \mathcal{O}_{j,j+1}=e^{i(n_F^0m-\Phi)(2j+1) x}\exp(\varphi_j-\varphi_{j+1}+m(\Theta_j+\Theta_{j+1})).
\end{equation}

At filling fractions $\nu=n_F^0/\Phi=1/m$, the phase of this operator cancels. For appropriate
interactions, the operator $\mathcal{O}_{j,j+1}$ becomes relevant under RG flow.
In this case, all the modes but the last one are gapped,  and
the system becomes topologically equivalent to a quantum Hall annulus. This results in 
the single  chiral state at the circumference  of the disc. 
The similar construction for TIs  takes into account the degeneracy between Kramer's partners, and is done
using a set of concentric wires hosting spinfull electrons.

\subsection{Time Reversal Invariance}\label{ap:TR}

Time reversal is implemented by an antiunitary operator $\mathcal{T}$ with $\mathcal{T}^2=-1$ for spin 1/2 particles.
We use the representation 

\begin{equation}\label{time_rev}
\mathcal{T}=i\tau_x\sigma_y\mathcal{K},
\end{equation}

\noindent where $\tau_x$ acts in the chirality index as $\tau_x\psi_{\eta,s}=\psi_{-\eta,s}$ and
$\sigma_y$ Pauli matrix acts on the spin index by $\sigma_y\psi_{\eta,s}=(\sigma_y)_{s,s'}\psi_{\eta,s'}$.
$\mathcal{K}$ represents complex conjugation.

For (1+1)-dimensional topological insulators, we can study the classification of time reversal
systems by examining the representative Dirac Hamiltonian for spin 1/2 particles

\begin{equation}\label{Dirac_ham}
 H=\int dx \bar{\psi}_s(-i\gamma^1\partial_x+m)\psi_s.
\end{equation}

\noindent with $\bar{\psi}=\psi^\dagger\gamma_0$. The two component spinor $\psi_s$ contains the chiral fields for 
each spin $s$. The matrices $\gamma^1$ and $\gamma^0$ satisfy $\{\gamma^\mu,\gamma^\nu\}=g^{\mu\nu}$. We use the 
chiral representation

\begin{equation}
  \gamma^0=\left(
\begin{array}{cc}
0 & 1 \\
1 & 0 
\end{array}
\right),\quad 
 \gamma^1=\left(
\begin{array}{cc}
0 & 1 \\
-1 & 0 
\end{array}
\right).
\end{equation}

Introducing a regularization scheme (studying the system on a $N$-site lattice for example) the Hamiltonian (in first quantization)
becomes a $2N\times 2N$ block diagonal matrix, where each
$N\times N$ block acts nontrivially in the chirality subspace.
The time reversal invariance of (\ref{Dirac_ham}) is reflected in

\begin{equation}\label{Symplectic}
 \mathcal{T}H\mathcal{T}^{-1}=H\rightarrow (i\sigma_y)H^T(-i\sigma_y)=H,
\end{equation}

\noindent where we have used the hermiticity of the Hamiltonian. Note that the last equality in (\ref{Symplectic})
is the definition of the symplectic Lie algebra $\mathfrak{sp}(2N)$. Hamiltonians satisfying (\ref{Symplectic}) are said to 
belong to the AII or symplectic class. The evolution operator $\exp(iHt)$ is then an element of the symmetric space 
(coset) $U(2N)/Sp(2N)$.

If we bosonize the fermionic Hamiltonian (\ref{Dirac_ham}), we obtain 

\begin{equation}\label{Ham_boson}
 H=\frac{1}{2\pi}\sum_{s=-}^+\int dx (\partial_x\theta_s)^2+(\partial_x\varphi_s)^2+\frac{m}{2}\cos(2\theta_s)
\end{equation}

Using (\ref{time_rev}) and (\ref{bosonization}) we find that the bosonic fields transform under time reversal as
(here $\bar{s}=-s$)

\begin{eqnarray}\label{time_rev_or}
 \mathcal{T}\theta_s\mathcal{T}^{-1}=\theta_{\bar{s}},\quad \mbox{and}\quad
 \mathcal{T}\varphi_s\mathcal{T}^{-1}=-\varphi_{\bar{s}}+\frac{1-s}{2}\pi.
\end{eqnarray}

\noindent so the Hamiltonian (\ref{Ham_boson}) written in terms of bosonic fields is invariant under TR (as it should be!).
Note that although (\ref{Ham_boson}) is a non-quadratic Hamiltonian in the bosonic fields, still we can associate to it a 
symmetric space, inherited from the fermionic description.

\subsubsection{Special points in parameter space}

It is worth noticing that some regions of the parameter space in the interacting Hamiltonian (\ref{Ham_micro}) 
correspond to a non interacting system in terms of link fields. In particular the region parametrized by
$\alpha_{\bar{\varphi}\varphi}=\alpha_{\varphi\theta}=\alpha_{\theta\bar{\theta}}=0$ and

\begin{eqnarray}\label{special_point}
\frac{(\alpha_{\theta\theta}\alpha_{\varphi\varphi}-\alpha_{\varphi\bar{\theta}}^2)}{\alpha_{\varphi\varphi}^2m^2}=K_\rho^{-2}=K_\sigma^{-2}=K^{-2},\quad
4\pi^2m^2(\alpha_{\theta\theta}\alpha_{\varphi\varphi}-\alpha_{\varphi\bar{\theta}}^2)=u_\rho=u_\sigma=u.
\end{eqnarray}

\noindent corresponds to two independent Luttinger liquids with the same Luttinger parameter $K$. At $K=1$, the
link fields can be mapped to non interacting fermions.
 
 \subsection{Renormalization of Tunneling operators}\label{sec:RG}
 
 The RG equations (\ref{RG_eqs}) can be linearized around the fixed point $g=0$ and $\Delta=2$. This last point
 defines a surface in parameter space 
 
 \begin{equation}\label{surface_app}
\left(\frac{K_\rho+K_\sigma}{4}\right)^2=1+\frac{|c|^2x}{(1+x)^2}, 
\end{equation}

\noindent for the Luttinger liquid parameters and velocities ($x\equiv u_\rho/u_\sigma$). Around the point $\mathbf{p}^0=(K^0_\rho,K_\sigma^0,x^0)$ lying in 
the surface (\ref{surface_app}), we expand $\Delta$

\begin{equation}
 \Delta(\mathbf{p}+\delta\mathbf{p})\approx 2+\frac{2}{K^0_\rho+K_\sigma^0}(\delta K_\rho+\delta K_\sigma)-\frac{|c|^2(1-x^0)}{(1+x^0)((1+x^0)^2+|c|^2x^0)}\delta x
\end{equation}
\noindent where $\delta \mathbf{p}=(\delta K_\rho,\delta K_\sigma,\delta x)$ and $|\delta \mathbf{p}|\ll 1$. The RG equations
(\ref{RG_eqs}) linearized around the point $\mathbf{p}$ and $g=0$ become

\begin{eqnarray}
\frac{dg}{dl}&=&-\left[\frac{2}{K^0_\rho+K_\sigma^0}(\delta K_\rho+\delta K_\sigma)-\frac{|c|^2(1-x^0)}{(1+x^0)((1+x^0)^2+|c|^2x^0)}\delta x\right]g,\\
\frac{d}{dl}(\delta K_\rho)&=&-\left[f^0(1+(f^0)^2)(K_\rho^0)^2\right]\left(\frac{g}{u_\rho^0}\right)^2,\\
\frac{d}{dl}(\delta K_\sigma)&=&-\left[f^0\frac{((x^0)^2+(f^0)^2)}{x^0}(K_\sigma^0)^2\right]\left(\frac{g}{u_\rho^0}\right)^2,\\
\frac{d}{dl}(\delta x)&=&\frac{1}{u^0_\sigma}\frac{d}{dl}(\delta u_\rho)-\frac{u_\rho^0}{(u^0_\sigma)^2}\frac{d}{dl}(\delta u_\sigma)=[K_\rho^0x^0(1-(f^0)^2)-K_\sigma^0((x^0)^2-(f^0)^2)]f^0\left(\frac{g}{u_\rho^0}\right)^2.
\end{eqnarray}

\noindent where $f^0=f(x^0)$, defined in (\ref{f_function}). Defining the variable $\lambda=\frac{2}{K^0_\rho+K_\sigma^0}(\delta K_\rho+\delta K_\sigma)-\frac{|c|^2(1-x^0)}{(1+x^0)((1+x^0)^2+|c|^2x^0)}\delta x$
we can study the renormalization of $\lambda$ and $y=g/u_\rho^0$. Combining the last three equations we get the BKT type of 
RG equations

\begin{eqnarray}
 \frac{dy}{dl}&=&-\lambda y \quad \mbox{and}\quad  \frac{d\lambda}{dl}=-\mathcal{C}(K_\rho^0,K_\sigma^0,u_\rho^0,u_\sigma^0)y^2
\end{eqnarray}

\noindent controlled by the sign of the function $\mathcal{C}$. Defining the always positive function

\begin{equation}
A=\frac{2f^0((1+(f^0)^2)(K^0_\rho)^2+((x^0)^2+(f^0)^2)(K_\sigma^0)^2/x^0)}{K_\rho^0+K_\sigma^0},
\end{equation}

\noindent $\mathcal{C}$ is explicitly

\begin{equation}\label{f_RG}
 \mathcal{C}=A\left(1+\frac{4|c|^2(1-x^0)x^0}{(1+x^0)^3(K_\rho^0+K_\sigma^0)}\left[\frac{(1-f^2)K^0_\rho+(f^2-(x^0)^2)K_\sigma^0/x^0}{(1+f^2)(K^0_\rho)^2+(x^0+f^2/{x^0})(K_\sigma^0)^2}\right]\right)
\end{equation}

\noindent where the values $(K_\rho^0,K_\sigma^0,x^0)$ are related by the equation $\Delta(K_\rho^0,K_\sigma^0,x^0)=2$. Note that
for $x\ll 1, x\gg1$ and $x=1$ $\mathcal{C}$ is positive.

 \subsection{Discretized Chern-Simons and continuous limit}\label{app:CS}
 
 The actions (\ref{CS_discrete}) and (\ref{ch_anomalies}) differ in the bulk by the term $\frac{1}{4\pi|m|}\int d^3x\left(A_x\partial_y A_\tau-A_\tau\partial_y A_x\right)$.
 This term appears disguised in the wire construction. It is related to gapless edge modes in the $x$ direction 
 in the construction based on Luttinger liquids, for a finite size system. These gapless edge modes in the $x$ 
 direction are inevitably connected to the seemingly missing term $\frac{1}{4\pi|m|}\int d^3x\left(A_x\partial_y A_\tau-A_\tau\partial_y A_x\right)$,
 as is shown below.
 
 Let us recall how the bulk Chern-Simons term is related to the edge modes as a consequence of gauge invariance \cite{Elitzur1989,WenBook2004}. The 
 action for the hydrodynamic field $b^+_\mu$ is given by (\ref{eff_action})
 
 \begin{equation}\label{CS_app}
  CS^+[b]=\frac{|m|}{4\pi}\int d^3x\epsilon^{\mu\nu\rho}b^+_\mu\partial_\nu b^+_{\rho}
 \end{equation}

 \noindent where we take the effective layer $+$ for simplicity. Under a gauge transformation $[b^+_\mu]^G=b^+_\mu-\partial_\mu\alpha$,
 the gauge transformed action $CS^+[b]$ becomes
 
\begin{eqnarray}\nonumber\label{CS_app_gauge}
  [CS^+[b]]^G&=&\frac{|m|}{4\pi}\int_{\Sigma} d^3x\epsilon^{\mu\nu\rho}[b^+_\mu]^G\partial_\nu [b^+_{\rho}]^G=\frac{|m|}{4\pi}\int_\Sigma d^3x\epsilon^{\mu\nu\rho}(b^+_\mu-\partial_\mu\alpha)\partial_\nu b^+_{\rho},\\
  \mbox{i.e}\quad [CS^+[b]]^G&=& CS^+[b]-\frac{|m|}{4\pi}\int_{\Sigma} d^3x\epsilon^{\mu\nu\rho}\partial_\mu\alpha\partial_\nu b^+_{\rho},
  \end{eqnarray}
 
 \noindent where we have used that $\epsilon^{\mu\nu\rho}\partial_\nu\partial_\rho\alpha=0$. The last term in (\ref{CS_app_gauge})
 can be integrated by parts in the manifold $\Sigma=\mathbb{R}\times\{[0,L_x]\times[0,L_y]\}$, leading to 

 \begin{equation}\label{CS_gt}
  [CS^+[b]]^G-CS^+[b]=\frac{|m|}{4\pi}\left\{\int_{-\infty}^\infty d\tau\int_0^{L_y} dy\alpha\left.(\partial_y b^+_\tau-i\partial_\tau b^+_y)\right\lvert_{x=0}^{x=L_x}-\int_{-\infty}^\infty d\tau\int_0^{L_x} dx\alpha\left.(\partial_x b^+_\tau-i\partial_\tau b^+_x)\right\lvert_{y=0}^{y=L_y}\right\}.
 \end{equation}
  
 \noindent assuming that the gauge transformation $\alpha$
 vanishes in the infinite past and infinite future. It is evident that the Chern-Simons action alone is not invariant under 
 a gauge transformation in a manifold with boundary. One way to solve this problem (making (\ref{CS_gt}) vanish identically) is to impose that the gauge field $b^+_\mu$
 becomes a pure gauge on the boundary, i.e. $b_\mu|\partial\Sigma=\partial_\mu\phi$. This resolution makes the fields
 at the boundary dynamical as (\ref{CS_app}) becomes
 
 \begin{eqnarray}
  CS^+[b]&=&\frac{|m|}{4\pi}\int_{-\infty}^{\infty}d\tau\int_{0}^{L_x} dx \int_0^{L_y} dy \left\{b^+_\tau(\partial_{x}b^+_{y}-\partial_{y}b^+_{x})+
  b^+_x(\partial_{y}b^+_{\tau}-i\partial_{\tau}b^+_{y})+b^+_y(i\partial_{\tau}b^+_{x}-\partial_{x}b^+_{\tau})\right\},\\
  &=&\frac{|m|}{2\pi}\int_{-\infty}^\infty d\tau\int_0^{L_x} dx\int_0^{L_y} dy\left\{ b^+_\tau(\partial_x b^+_y-\partial_yb^+_x)+\frac{b^+_y(i\partial_\tau b^+_x)-b^+_x(i\partial_\tau b^+_y)}{2}\right\}\\
  &+&\frac{|m|}{4\pi}\left\{\int_{-\infty}^\infty d\tau\int_0^{L_x} dx\left.(b^+_x b^+_\tau)\right\lvert_{y=0}^{y=L_y}-\int_{-\infty}^\infty d\tau\int_0^{L_y} dy\left.(b^+_y b^+_\tau)\right\lvert_{x=0}^{x=L_x}\right\}.\label{boundary}
 \end{eqnarray}

 Using the condition of pure gauge at the boundary in (\ref{boundary}) we have
 
 \begin{eqnarray}\label{CS_bulk}
  CS^+[b]&=&\frac{|m|}{2\pi}\int_\Sigma\left\{ b^+_\tau(\partial_x b^+_y-\partial_yb^+_x)+\frac{b^+_y(i\partial_\tau b^+_x)-b^+_x(i\partial_\tau b^+_y)}{2}\right\}\\
  &+&\frac{|m|}{4\pi}\left\{\int_{-\infty}^\infty d\tau\int_0^{L_x} dx\left.(\partial_x\phi^+)( i\partial_\tau\phi^+)\right\lvert_{y=0}^{y=L_y}-\int_{-\infty}^\infty d\tau\int_0^{L_y} dy\left.(\partial_y\phi^+)(i\partial_\tau\phi^+)\right\lvert_{x=0}^{x=L_x}\right\},
 \end{eqnarray}

 \noindent note that the boundary term $(\partial_x\phi^+)( i\partial_\tau\phi^+)\lvert_{y=0}^{y=L_y}$ appears
 integrating by parts $b_x\partial_yb_\tau-b_\tau\partial_yb_x$. In the partition function $\mathcal{Z}=\int\mathcal{D}be^{-CS[b]}$ the hydrodynamic field $b_\mu^+$ is integrated over. The path
 integral over $b_\tau$ imposes the constraint $(\partial_x b^+_y-\partial_yb^+_x)=0$ which makes the fields $b^+_x,b^+_y$
 pure gauge in the whole manifold. This implies that the r.h.s of (\ref{CS_bulk}) vanishes. The Chern-Simons action
 depends just on the fields in the boundary of the manifold
 
 \begin{eqnarray}\label{CS_boundary}
  CS^+[b]&=&\frac{|m|}{4\pi}\left\{\int_{-\infty}^\infty d\tau\int_0^{L_x} dx\left.(\partial_x\phi^+)( i\partial_\tau\phi^+)\right\lvert_{y=0}^{y=L_y}-\int_{-\infty}^\infty d\tau\int_0^{L_y} dy\left.(\partial_y\phi^+)(i\partial_\tau\phi^+)\right\lvert_{x=0}^{x=L_x}\right\},
 \end{eqnarray}
 
 Finally, to connect with the wire construction, we take $L_x\rightarrow\infty$, obtaining
 
  \begin{eqnarray}
  CS^+[b]=\frac{|m|}{4\pi}\int_{-\infty}^\infty d\tau\int_{-\infty}^{\infty} dx\left[(\partial_x\phi^+_L)( i\partial_\tau\phi^+_L)-(\partial_x\phi^+_0)( i\partial_\tau\phi^+_0)\right],
 \end{eqnarray}
  
  \noindent which describes the chiral edge modes in the first and last wires (for the $+$ layer). As we discussed, 
  these edge modes appear as boundary terms after integrating  $\frac{|m|}{4\pi}\int_\Sigma (b^+_x\partial_yb^+_\tau-b^+_\tau\partial_yb^+_x)$ by parts. 
  Coupling the hydrodynamic field $b^+_\mu$ to an external gauge field $A_\mu$ and integrating over $b^+_\mu$, generates 
  the term $\frac{1}{4\pi|m|}\int_\Sigma (A_x\partial_yA_\tau-A_\tau\partial_yA_x)$.

  Including the action from both layers, and using $L_y=dN$, with $d$ the distance between the wires, we have
  
  \begin{equation}
  CS^+[b]-CS^-[b]=\frac{|m|}{4\pi}\sum_{s=+,-}s\int_{-\infty}^\infty d\tau\int_{-\infty}^{\infty} dx\left[(\partial_x\phi^s_N)( i\partial_\tau\phi^s_N)-(\partial_x\phi^s_0)(i\partial_\tau\phi^s_0)\right].
  \end{equation}
  
  This action together with the Hamiltonian describing the gapless edges (\ref{edge_ham}) amounts to the full description of the
  edge dynamics
  
  \begin{equation}
   S_{\rm edge}=\frac{|m|}{4\pi}\sum_{s=+,-}\int_{-\infty}^\infty d\tau\int_{-\infty}^{\infty} dx\left[(\partial_x\phi^s_N)( si\partial_\tau\phi^s_N+v_F\partial_x\phi^s_N)+(\partial_x\phi^s_0)(-si\partial_\tau\phi^s_0+v_F\partial_x\phi^s_0)\right].
  \end{equation}

 
 
\end{document}